\documentclass[twocolumn,prd,aps,nofootinbib,showpacs]{revtex4}
\usepackage{multirow}
\usepackage{graphicx}
\usepackage[dvips]{color}
\usepackage{amssymb,amsmath}

\begin{document}

\title{The preliminary lattice QCD calculation of $\kappa$ meson decay width}

\author{Ziwen Fu}
\affiliation{
Key Laboratory of Radiation Physics and Technology {\rm (Sichuan University)},
Ministry of Education; \\
Institute of Nuclear Science and Technology, Sichuan University,
Chengdu 610064, P. R. China.
}

\begin{abstract}
We present a direct lattice QCD calculation of
the $\kappa$ meson decay width with the $s$-wave scattering phase shift
for the isospin $I=1/2$ pion-kaon ($\pi K$) system.
We employ a special finite size formula,
which is the extension of the Rummukainen-Gottlieb formula
for the $\pi K$ system in the moving frame, to calculate 
the scattering phase, which indicates a resonance around $\kappa$ meson mass.
Through the effective range formula,
we extract the effective $\kappa \to \pi K$ coupling
constant $g_{\kappa \pi K} = 4.54(76)$ GeV
and decay width  $\Gamma = 293 \pm 101$~MeV.
Our simulations are done with the MILC gauge configurations
with $N_f=2+1$ flavors of the ``Asqtad'' improved staggered dynamical sea quarks
on a $16^3\times48$ lattice 
at $(m_\pi + m_K) / m_\kappa \approx 0.8$
and lattice spacing $a \approx 0.15$ fm.
\end{abstract}
\pacs{12.38.Gc, 11.15.Ha}

\maketitle

\section{ Introduction }
\label{Sec:Introduction}
It is well-known that kappa meson is a resonance,
which is a state with a considerable width under strong interactions,
and decay via strong interaction.
In 2010, Particle Data Group (PDG)~\cite{Nakamura:2010zzi}
lists the meson $K_0^*(800)$,
which is usually called $\kappa$ meson ($J^P=0^+$, $I=1/2$),
with a  mass ($676\pm40$ MeV) and a broad width ($548\pm24$ MeV).
Some recent experimental analyses~\cite{Bugg:2005xx,Bugg:2009uk,Aitala:2002kr,
Ablikim:2010kd, Ablikim:2005ni,Bai:2003fv,
Alde:1997ri,Markushin:1999fg,
Xiao:2000kx,Cawlfield:2006hm,Bonvicini:2008jw}
and phenomenological analysis of $\pi K$ scattering
~\cite{Ishida:1997wn,Zhou:2006wm,DescotesGenon:2006uk,
Bugg:2003kj,Zheng:2003rw,Pelaez:2004xp}
strongly show its existence.
The Dalitz plot analysis of Fermilab experiment E791~\cite{Aitala:2002kr}
yield its decay width about $410 \pm 43 \pm 87$~MeV.
Moreover, the $\kappa$ meson has been extensively studied
with BES data~\cite{Ablikim:2010kd,Ablikim:2005ni},
where the evidence for its existence is very clear,
and the most recent analyses  based on events collected by BESII
gives its mass about $849 \pm 77$~MeV, and decay width about $512 \pm 80$~MeV.

Although the direct determination of $\kappa$ resonance parameters
from QCD is extremely hard since the calculation of resonance masses
and decay widths is essentially a nonperturbative problem.
Several research groups have undertaken theoretical efforts to
study the $\kappa$ meson and its resonance parameters~\cite{Ishida:1997wn,Zhou:2006wm,DescotesGenon:2006uk,
Bugg:2003kj,Zheng:2003rw,Pelaez:2004xp}.
However, all the experimental and theoretical analyses
give a little bit different resonance mass and
decay width for $\kappa$ meson.
Therefore,  there is not close to the consensus
yet on its resonance parameters.

The feasible way to study $\kappa$ resonance nonperturbatively
from first principles is with the help of lattice QCD.
Until now, no direct lattice QCD study about $\kappa$ resonance
have been reported,
possibly because the reliable calculation of the rectangular diagram are difficult,
and there are not suitable theoretical formula available
to represent the $\pi K$ system in the moving frame.
Inspired by J.~Nebreda and J.~Pelaez's theoretical studies on
$\kappa$ resonance~\cite{Nebreda:2010wv}
and our previous works about precise extraction
of $\kappa$ mass~\cite{Fu:2011zz},
and reliable extraction of $\pi K$ scattering length~\cite{Fu:2011wc}
at $I=1/2$ channel,
in this work, we further explore its resonance parameters
directly from lattice QCD.

In this paper, we estimate the $\kappa$ meson decay width
by calculating the $s$-wave scattering phase shift
for the $I=1/2$ $\pi K$ system.
We discuss them both in center of mass (CM) frame and the moving frame.
The simulation are done with MILC lattice ensemble~\cite{Bernard:2010fr,Bazavov:2009bb}.
The meson masses determined in~\cite{Fu:2011zz}
gave $(m_\pi+m_K)/m_\kappa=0.8$~\cite{Fu:2011zz},
and the lattice parameters determined
by MILC collaboration are: the lattice extent $L \approx 2.5\ {\rm fm}$ 
and the lattice space inverse $1/a=1.358\ {\rm GeV}$~\cite{Bernard:2010fr,Bazavov:2009bb}.
The finite size formula~\cite{Luscher:1986pf,Luscher:1991cf,Lellouch:2001p4241,
Luscher:1991ux,Luscher:1990ck}
is employed to $\pi K$ system in the center-of-mass frame,
and we utilize a special finite size formula,
which is the extension of the Rummukainen-Gottlieb formula
for the $\pi K$ system in the moving frame~\cite{Fu:2011xz},  
to compute the scattering phase shift.

\section{Methods}
\label{Sec:Methods}
\subsection{The effective range formula }
\label{SubSec:Scattering_Phase}
In the $\pi K$ system, the relativistic Breit-Wigner formula (RBWF) for the elastic
$s$-wave scattering phase $\delta_0$ in the resonance region
with a center-of-mass energy $m_\kappa$
and a decay width $\Gamma_R$ can be conveniently expressed as
\begin{equation}
\label{eq:BW}
\tan\delta_0=\frac{\sqrt{s} \, \Gamma_R(s)}{m_\kappa^2-s} ,
\end{equation}
where $s=E_{CM}^2$ is the invariant mass of the $\pi K$ system,
and $E_{CM}$ is its center-of-mass energy.
The quantum numbers of $\kappa$ resonance is
$I(J^{P})=\frac{1}{2}(0^{+})$
and decays into one pion and one kaon in the $s$-wave, which can be handle on the lattice.
The decay width $\Gamma_R(s)$ can be written in terms of the
coupling constant $g_{\kappa \pi K}$~\cite{Nebreda:2010wv},
\begin{equation}
\label{eq:g_kpK_formula}
\Gamma_R(s)=\frac{g^2_{\kappa\pi K}}{8\pi}\frac{p}{s}\;,
\end{equation}
where
\begin{equation}
\label{eq:effective_range_formula2}
p = \frac{1}{2\sqrt{s}}
\sqrt{ \left[s - (m_\pi - m_K)^2\right]\left[ s - (m_\pi + m_K)^2\right] }\,,
\end{equation}
is the center-of-mass momentum of the pion or kaon.

With a combination of Eqs.~(\ref{eq:BW}) and (\ref{eq:g_kpK_formula}),
a description of the scattering phase in the $s$-wave
as a function of the invariant mass $\sqrt{s}$ is provided
by so-called the effective range formula (ERF) in the elastic region,
\begin{equation}
\label{eq:effective_range_formula}
\tan{\delta_0}=\frac{g^2_{\kappa\pi K}}{8\pi}
\frac{p}{ \sqrt{s}(m_R^2- s)} \,.
\end{equation}
This equation allows us a fit or
seeking for the desired parameter $g_{\kappa\pi K}$,
which is the effective $\kappa\rightarrow\pi K$ coupling constant,
and the resonance position $m_\kappa$.

The $\kappa$ decay width $\Gamma_\kappa$
can then be calculated through,
\begin{eqnarray}
\label{eq:decay_width}
\Gamma_\kappa \hspace{-0.2cm}&=&\hspace{-0.2cm}\Gamma_R(s)\Bigg|_{s=m_\kappa^2} =
\frac{g^2_{\kappa\pi K}}{8\pi}\frac{p_\kappa}{m_\kappa^2} \,, \\
p_\kappa \hspace{-0.2cm}&=&\hspace{-0.2cm}
\frac{1}{2m_\kappa}\sqrt{ [m_\kappa^2 - (m_\pi - m_K)^2][ m_\kappa^2 - (m_\pi+m_K)^2] } \;. \nonumber
\end{eqnarray}
Therefore, equations~(\ref{eq:effective_range_formula}) and (\ref{eq:decay_width})
allow us to archive $m_\kappa$ and $\Gamma_\kappa$
by way of the dependence of $\delta_0$ on $s$.

\subsection{Finite-volume method}

\subsubsection{Center of mass frame}
In the center-of-mass frame, if we doesn't consider
the interaction between pion and kaon,
the possible energy eigenvalues of $\pi K$ system reads
$$
E = \sqrt{m_\pi^2+ p^2} + \sqrt{m_K^2 + p^2} \,,
$$
where $p=|{\mathbf p}|, \; {\mathbf p}=(2\pi/L){\mathbf n}$, and
${\mathbf n}\in \mathbb{Z}^3$.
In a typical lattice study, the energy for the ${\mathbf n} \ne 0$
is larger than kappa mass $m_\kappa$.
For instance, on our chosen gauge configurations,
the lowest energy for the ${\mathbf n} \ne 0$  evaluated from $m_\pi$, $m_K$ and $m_\kappa$~\cite{Fu:2011zz},
is $E = 1.35\times m_\kappa$,
which is not eligible to study the $\kappa$ meson decay.
So, we must consider the ${\mathbf n} = 0$ case.
On our chosen gauge configurations the invariant mass of free pion and kaon
takes $\sqrt{s} = 0.8 \times m_\kappa$,
which is away from kappa mass $m_\kappa$,
but closer to $m_\kappa$ than those with the $ {\mathbf n} \ne 0$.

With the consideration of the interaction of $\pi K$ system.
The energy eigenvalues are moved from $E$ to $\overline{E}$,
and the energy eigenvalue for $\pi K$ system is,
$$
\overline{E} = \sqrt{m_\pi^2 + k^2} + \sqrt{m_K^2 + k^2} ,
\quad k=\frac{2\pi}{L}q ,
$$
where $q \in \mathbb{R}$. From this equation, we obtain the scattering momentum $k$
\begin{equation}
\label{eq:CM_k}
k = \frac{1}{2\overline{E}}\sqrt{ [ \overline{E}^2 \hspace{-0.03cm}-\hspace{-0.03cm}(m_\pi \hspace{-0.03cm}-\hspace{-0.03cm} m_K)^2][ \overline{E}^2 \hspace{-0.03cm}-\hspace{-0.03cm} (m_\pi\hspace{-0.03cm}+\hspace{-0.03cm}m_K)^2] }\;.
\end{equation}

In the center-of-mass frame these energy eigenstates transform
under the cubic group $O(3)$
in the irreducible representation $\Gamma = T_1^+$.
The finite-size formula
connecting the energy $\overline{E}$ or $\overline{E}$
to the $s$-wave $\pi K$ scattering phase shift $\delta_0$ 
is~\cite{Luscher:1986pf,Luscher:1991cf,
Lellouch:2001p4241,Luscher:1991ux,Luscher:1990ck}
\begin{equation}
\label{eq:CMF}
\tan\delta_0(k)=\frac{\pi^{3/2}q}{\mathcal{Z}_{00}(1;q^2)} ,
\end{equation}
where the zeta function is formally denoted by
\begin{equation}
\label{eq:Zeta00_CM}
\mathcal{Z}_{00}(s;q^2)=\frac{1}{\sqrt{4\pi}}
\sum_{{\mathbf n}\in\mathbb{Z}^3} \frac{1}{\left(|{\mathbf n}|^2-q^2\right)^s} .
\end{equation}

\subsubsection{Moving frame}
In order to make the energy of $\pi K$ system is
more close to the $\kappa$ mass $m_\kappa$,
we consider the moving frame (MF)~\cite{Rummukainen:1995vs}.
Using a moving frame with total momentum ${\mathbf P}=(2\pi/L){\mathbf d}$,
${\mathbf d}\in\mathbb{Z}^3$,
the energy eigenvalues for non-interacting $\pi K$ system are
\begin{equation}
E_{MF} = \sqrt{m_\pi^2+p_1^2} + \sqrt{m_K^2+ p_2^2}\,,
\end{equation}
where $p_1=|{\mathbf p}_1|$,
      $p_2=|{\mathbf p}_2|$,
and ${\mathbf p}_1$, ${\mathbf p}_2$ denote
the three-momenta of the pion and kaon meson, respectively,
which satisfy the periodic boundary condition,
$$
{\mathbf p}_i=\frac{2\pi}{L}{\mathbf n}_i\,,
\quad {\mathbf n}_i\in \mathbb{Z}^3\,,
$$
and the relation
$
\quad{\mathbf p}_1+{\mathbf p}_2={\mathbf P}\,.
$

Using the standard Lorentz transformation
with a boost factor $\gamma=1/\sqrt{1-{\mathbf v}^2}$, 
here ${\mathbf v}={\mathbf P}/E_{MF}$.
The $E_{CM}$ can be obtained through
\begin{equation}
E_{CM} = \gamma^{-1}E_{MF} = \sqrt{m_\pi^2 + p^{*2}} + \sqrt{m_K^2+ p^{*2}} \,,
\end{equation}
where, in the center-of-mass frame, the total center-of-mass momentum disappears, 
$$
p^*=| {\mathbf p}^*|\,,\quad
{\mathbf p}^*={\mathbf p}^*_1=-{\mathbf p}^*_2\,,
$$
here and later we denote the center-of-mass momenta with an asterisk $(^\ast)$.
The boost factor acts in the direction of the velocity ${\mathbf v}$,
here we use the shorthand notation
\begin{equation}
\label{la:gamma}
\vec{\gamma}{\mathbf p} =
\gamma{\mathbf p}_{\parallel}+{\mathbf p}_{\perp}\,,\quad
\vec{\gamma}^{-1}{\mathbf p} =
\gamma^{-1}{\mathbf p}_{\parallel}+{\mathbf p}_{\perp}\,,
\end{equation}
where ${\mathbf p}_{\parallel}$ and ${\mathbf p}_{\perp}$ are
the components of ${\mathbf p}$ parallel and perpendicular to
the center-of-mass velocity, respectively, 
\begin{equation}
{\mathbf p}_{\parallel}=
\frac{{\mathbf p}\cdot{\mathbf v}}{|\mathbf v|^2}{\mathbf v}\,,\quad
{\mathbf p}_{\perp}={\mathbf p}-{\mathbf p}_{\parallel}\,.
\end{equation}
The ${\mathbf p}^*$ are quantized to the values~\cite{Fu:2011xz}
\begin{equation}
{\mathbf p}^* =\frac{2\pi}{L}{\mathbf r}\,, \qquad
{\mathbf r} \in P_{\mathbf d}\,,
\end{equation}
where the set $P_{\mathbf d}$  is
\begin{equation}
\label{eq:set_Pd_MF}
P_{\mathbf d} = \left\{ {\mathbf r} \left|  {\mathbf r} = \vec{\gamma}^{-1}
\left[{\mathbf n}+\frac{{\mathbf d}}{2}
\left(1+\frac{m_K^2\hspace{-0.1cm}-\hspace{-0.1cm}m_\pi^2}{E_{CM}^2}\right)
\right], \right. \ {\mathbf n}\in\mathbb{Z}^3 \right\} \,,
\end{equation}

Here we only consider the dominant low energy states
in the moving frame: the pion with zero momentum,
and kaon with the momentum
${\mathbf p} = (2\pi / L) {\mathbf e}_3$ (${\mathbf d} = {\mathbf e}_3$)
and the $\kappa$ meson with the momentum  ${\mathbf P} = {\mathbf p}$.
For our case, the invariant mass of $\pi K$ system
has $\sqrt{s} = 0.93\times m_\kappa$,
which is much closer to $\kappa$ mass $m_\kappa$
than these in the center-of-mass frame.
So, we only investigate this case.
The invariant mass $\sqrt{s}$ was calculated by
$\sqrt{s} = \sqrt{ E_{MF}^2 - P^2 }$.

In the interacting case, the $\overline{E}_{CM}$ is provided by
\begin{equation}
\label{eq:continuum_dis_rel}
\overline{E}_{CM} = \sqrt{m_\pi^2 + k^{2}} + \sqrt{m_K^2 + k^{2}}\,,
\quad k = \frac{2\pi}{L} q \,.
\end{equation}
where $q\in\mathbb{R}$. From this equation, we gain 
\begin{equation}
\label{eq:MF_k}
k = \frac{1}{2\overline{E}}\sqrt{ [ \overline{E}^2 \hspace{-0.03cm}-\hspace{-0.03cm}(m_\pi \hspace{-0.03cm}-\hspace{-0.03cm} m_K)^2][ \overline{E}^2 \hspace{-0.03cm}-\hspace{-0.03cm} (m_\pi\hspace{-0.03cm}+\hspace{-0.03cm}m_K)^2] }\;.
\end{equation}
An immediate consequence of Eq.~(\ref{eq:MF_k}) is
that the physical kinematics for the $\kappa$-meson decay satisfy
\begin{equation}
\overline{E}_{CM} \ge m_\pi + m_K ,
\end{equation}
namely, $\kappa$ can decay only if its mass exceeds
the sum of the masses of its decay products (namely, $\pi$ and $K$).
We can rewrite equqtion~(\ref{eq:MF_k}) to an elegant form for later use 
\begin{equation}
\label{eq:MF_k_e}
k^2  = \frac{1}{4}
\left( \overline{E}_{CM} + \frac{m_\pi^2 - m_K^2}{\overline{E}_{CM}} \right)^2 - m_\pi^2 \,.
\end{equation}

The energy eigenstates transform under the tetragonal group $C_{4v}$.
the irreducible representations of $A_1$  and $E$ are
relevant for the $s$-wave $\pi K$ scattering states in a cubic box.
In this work, we only consider the energies linked with the $A_1$ sector.
The hadron interaction moves the energy from $E$ to $\overline{E}$,
and the energies $\overline{E}$ are
associated to the $\pi K$ scattering phase shift
$\delta_0$ in a torus through
the $\pi K$ system's Rummukainen-Gottlieb formula~\cite{Rummukainen:1995vs},
which is an extension of the L\"uscher formula~\cite{Luscher:1986pf}
to the moving frame~\cite{Fu:2011xz},
\begin{equation}
\label{eq:Luscher_MF}
\tan\delta_0(k)=
\frac{\gamma\pi^{3/2}q}{\mathcal{Z}_{00}^{\mathbf d}(1;q^2)}\;,
\end{equation}
where the modified zeta function is formally defined by
\begin{equation}
Z_{00}^{{\mathbf d}} (s; q^2) = \sum_{{\mathbf r}\in P_{\rm d}}
\frac{1} { ( |{\mathbf r}|^2 - q^2)^s } \,,
\label{zetafunction_MF}
\end{equation}
and the set $P_{\mathbf d}$  is denoted in Eq.~(\ref{eq:set_Pd_MF}).
Equation~(\ref{zetafunction_MF}) is first provided in Ref.~\cite{Davoudi:2011md}
for generic two-particles system,
and we further confirmed and rigorously proved it~\cite{Fu:2011xz}.
The $k$ is the momentum defined from the invariant mass $\sqrt{s}$ as
$\sqrt{s} = \sqrt{ k^2 + m_\pi^2 } + \sqrt{ k^2 + m_K^2 }$.
The calculation method of $Z_{00}^{{\mathbf d}} (1; q^2)$
is full discussed in Appendix~\ref{appe:zeta}.

\subsection{Correlation matrix}
\label{SubSec:Correlation_matrix }
In order to calculate the two energy eigenstates,
namely, $\overline{E}_n$ ($n=1,2$),
we construct a matrix of the time correlation function,
\begin{equation}
C(t) = \left(
\begin{array}{ll}
\langle 0 | {\cal O}_{\pi K}^\dag(t)  {\cal O}_{\pi K}(0) | 0 \rangle &
\langle 0 | {\cal O}_{\pi K}^\dag(t)  {\cal O}_\kappa(0)  | 0 \rangle
\vspace{0.3cm}\\
\langle 0 | {\cal O}_\kappa^\dag(t)   {\cal O}_{\pi K}(0) | 0 \rangle &
\langle 0 | {\cal O}_\kappa^\dag(t)   {\cal O}_\kappa(0)  | 0 \rangle
\end{array} \right) ,
\label{eq:CorrMat}
\end{equation}
where ${\cal O}_\kappa(t)$ is an interpolating operator for the $\kappa$ meson
with the definite momentum, and ${\cal O}_{\pi K}(t)$ is
an interpolating operator for the $\pi K$ system with specified momentum.
The interpolating operators ${\cal O}_\kappa(t)$ and ${\cal O}_{\pi K}(t)$
employed here is exactly the same as in our previous studies~{\cite{Fu:2011wc,Fu:2011zz}.
To make this paper self-contained,
we will provide all the necessary definitions below.

\subsubsection{$\pi K$ sector}
Here we briefly review the required formulae to compute 
the scattering phase for the $\pi K$ system in a torus
using the original derivation and notation
in Refs.~\cite{Nagata:2008wk,Sharpe:1992pp,
Kuramashi:1993ka,Fukugita:1994na,Fukugita:1994ve}.
Let us study the $\pi K$ scattering system of one Nambu-Goldstone pion and
one Nambu-Goldstone kaon in the Asqtad-improved staggered dynamical fermion formalism.
Using the operators ${\cal O}_\pi(x_1), {\cal O}_\pi(x_3)$ for pions at points $x_1, x_3$,
and the operators ${\cal O}_K(x_2), {\cal O}_K(x_4)$ for kaons at points $x_2, x_4$, respectively,
with the pion and kaon interpolating field operators defined by
\begin{eqnarray}
{\cal O}_{\pi^+}({\mathbf{x}},t) \hspace{-0.1cm}&=&\hspace{-0.1cm}
- \overline{d}({\mathbf{x}},t)\gamma_5 u({\mathbf{x}},t) \nonumber \\
{\cal O}_{\pi^0}({\mathbf{x}},t) \hspace{-0.1cm}&=&\hspace{-0.1cm}
\frac{1}{\sqrt{2}}
[\overline{u}({\mathbf{x}},t)\gamma_5 u({\mathbf{x}},t) -
 \overline{d}({\mathbf{x}},t)\gamma_5 d({\mathbf{x}},t) ] , \nonumber \\
{\cal O}_{K^0}({\mathbf{x}},t)   \hspace{-0.1cm}&=&\hspace{-0.1cm}
\overline{s}({\mathbf{x}},t)\gamma_5 d({\mathbf{x}}, t), \nonumber \\
{\cal O}_{K^+}({\mathbf{x}},t)   \hspace{-0.1cm}&=&\hspace{-0.1cm}
 \overline{s}({\mathbf{x}},t)\gamma_5 u({\mathbf{x}},t) .
 \end{eqnarray}
We then represent $\pi K$ four-point functions as
$$
C_{\pi K}(x_4,x_3,x_2,x_1) =
\bigl< {\cal O}_K(x_4) {\cal O}_{\pi}(x_3) {\cal O}_K^{\dag}(x_2) {\cal O}_{\pi}^{\dag}(x_1)\bigr>.
$$
In this paper we only consider one moving frame, 
which is the pion with zero momentum,
and kaon with momentum
${\mathbf p} = (2\pi / L) {\mathbf e}_3$ (namely, ${\mathbf d}={\mathbf e}_3$)
and total momentum ${\mathbf P}={ \mathbf p}$.
The operator which creates a single kaon with non-zero three momentum ${\mathbf k}$
from the vacuum is obtained by Fourier transform:
$
{\cal O}_{K  }({\mathbf{k}}, t) = \sum_{\mathbf x} {\cal O}_{K}({\mathbf{x}},t)
e^{ i{\mathbf k} \cdot \mathbf{x} } \,.
$

After summing over the spatial coordinates
${\mathbf{x}}_1$, ${\mathbf{x}}_2$, ${\mathbf{x}}_3$ and ${\mathbf{x}}_4$,
we get the $\pi K$ four-point function 
\begin{eqnarray}
\label{EQ:4point_pK_mom}
C_{\pi K}({\mathbf p}, t_4,t_3,t_2,t_1) &=& \nonumber \\
&&\hspace{-2.5cm} \sum_{\mathbf{x}_1, \mathbf{x}_2, \mathbf{x}_3, \mathbf{x}_4}
e^{ i{\mathbf p} \cdot ({\mathbf{x}}_4 -{\mathbf{x}}_2) }
C_{\pi K}(x_4,x_3,x_2,x_1) \,,
\end{eqnarray}
in the momentum ${\mathbf p}$ state, 
where $x_1 \equiv ({\mathbf{x}}_1,t_1)$,  $x_2 \equiv ({\mathbf{x}}_2,t_2)$,
$x_3 \equiv ({\mathbf{x}}_3,t_3)$, and $x_4 \equiv ({\mathbf{x}}_4,t_4)$, and
$t$ stands for the time difference, namely, $t\equiv t_3 - t_1$.
To prevent the complicated Fierz rearrangement of the quark lines~\cite{Fukugita:1994ve},
we select $t_1 =0, t_2=1, t_3=t$, and $t_4 = t+1$.
We construct $\pi K$ operators for the $I=1/2$  channel as
\begin{eqnarray}
\label{EQ:op_pipi}
{\cal O}_{\pi K}^{I=\frac{1}{2}}({\mathbf{p}}, t) &=& \frac{1}{\sqrt{3}}
\Bigl\{  \sqrt{2}\pi^+(t) K^0({\mathbf{p}}, t+1) -  \nonumber \\
&& \hspace{1.2cm} \pi^{0}(t) K^{+}({\mathbf{p}}, t+1)  \Bigl\} \,,
\end{eqnarray}
where ${\mathbf{p}}$ is total momentum  of $\pi K$ system.
This $\pi K$ operator has $(I,I_z)=(1/2, 1/2)$.

Supposing that the masses of the $u$ and $d$ quarks are degenerate,
only three diagrams contribute to
the $\pi K $ scattering amplitudes~\cite{Nagata:2008wk}.
These quark line diagrams are shown in Fig.~\ref{fig:diagram},
which are labeled  as direct diagram (D), crossed diagram (C) and 
rectangular diagram (R), respectively.
The direct and crossed diagrams can be readily evaluated~\cite{Fukugita:1994ve}.
However, the rectangular diagram (R) needs another quark propagator
linking the time slices $t_3$ and $t_4$,
which make it hard to compute~\cite{Fukugita:1994ve}.

Sasaki et al. tackle this puzzle through the technique
with a fixed kaon sink operator~\cite{Sasaki:2010zz}.
We successfully extended the technique in Refs.~\cite{Kuramashi:1993ka,Fukugita:1994ve}
to the $\pi K$ scattering at the $I=1/2$ channel with zero momentum~\cite{Fu:2011wc},
here we further consider the case with the non-zero momentum.
To be specific, each propagator corresponding
to a moving wall source is given by~\cite{Kuramashi:1993ka,Fukugita:1994ve,Gupta:1990mr}
\begin{equation}
\sum_{n''}D_{n',n''}G_t(n'') = \sum_{\mathbf{x}}
\delta_{n',({\mathbf{x}},t)}, \quad 0 \leq t \leq T-1 .
\end{equation}

\begin{figure}[t]
\begin{center}
\includegraphics[width=8.0cm]{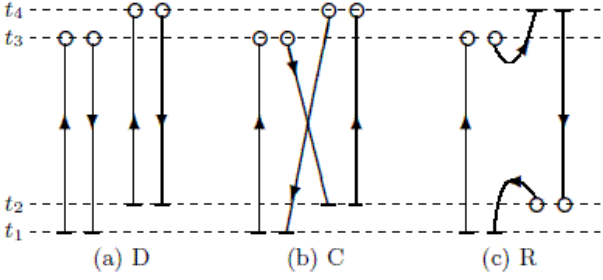}
\end{center}
\vspace{-0.5cm}
\caption{ \label{fig:diagram}
Diagrams contributing to $\pi K$ four-point functions.
Short bars stand for the wall sources.
Open circles are sinks for local pion or kaon operators.
The thicker lines represent the strange quark lines.
}
\end{figure}
For the non-zero momentum,  we used a up quark source with $1$, and a
strange quark source with $e^{i{\mathbf p} \cdot {\mathbf{x}} }$ on each site
for the pion and kaon  creation operator, respectively.
$D$, $C$, and $R$, are shown in Fig.~\ref{fig:diagram},
we can express them in terms of the quark propagators $G$, namely,
\begin{widetext}
\begin{eqnarray}
\label{eq:dcr}
C^D_{\pi K}({\mathbf p},t_4,t_3,t_2,t_1) &=&
\sum_{ {\mathbf{x}}_3}
\sum_{ {\mathbf{x}}_4} \, e^{ i{\mathbf p} \cdot {\mathbf{x}}_4 }
\left\langle \mbox{Re} \, \mbox{Tr}
[G_{t_1}^{\dag}({\mathbf{x}}_3, t_3) G_{t_1}({\mathbf{x}}_3, t_3)
 G_{t_2}^{\dag}({\mathbf{x}}_4, t_4) G_{t_2}({\mathbf{x}}_4, t_4) ] \right\rangle,\nonumber \\
 C^C_{\pi K}({\mathbf p},t_4,t_3,t_2,t_1) &=&
\sum_{ {\mathbf{x}}_3}
\sum_{ {\mathbf{x}}_4} \, e^{ i{\mathbf p} \cdot {\mathbf{x}}_4 }
\left\langle \mbox{Re} \, \mbox{Tr}
[G_{t_1}^{\dag}({\mathbf{x}}_3, t_3) G_{t_2}({\mathbf{x}}_3, t_3)
 G_{t_2}^{\dag}({\mathbf{x}}_4, t_4) G_{t_1}({\mathbf{x}}_4, t_4) ] \right\rangle,\nonumber \\
 C^R_{\pi K}({\mathbf p}, t_4,t_3,t_2,t_1) &=&
\sum_{ {\mathbf{x}}_2}
\sum_{ {\mathbf{x}}_3} \, e^{ i{\mathbf p} \cdot {\mathbf{x}}_2 }
\left\langle \mbox{Re} \, \mbox{Tr}
[G_{t_1}^{\dag}({\mathbf{x}}_2, t_2) G_{t_4}({\mathbf{x}}_2, t_2)
 G_{t_4}^{\dag}({\mathbf{x}}_3, t_3) G_{t_1}({\mathbf{x}}_3, t_3) ] \right\rangle .
\end{eqnarray}
\end{widetext}

All three diagrams in Figure~\ref{fig:diagram} are needed
to study $\pi K$ scattering in the $I=1/2$ channel.
If assuming that $u$ and $d$ quarks have the same mass,
the $\pi K$ correlation function for $I=1/2$ channel
can be expressed as the combinations of these three diagrams~\cite{Nagata:2008wk},
\begin{equation}
\label{EQ:phy_I12_32}
C_{\pi K}(t)  \equiv
\left\langle {\cal O}_{\pi K}(t) | {\cal O}_{\pi K}(0) \right\rangle =
D + \frac{1}{2} C - \frac{3}{2} R ,
\end{equation}
where the operator ${\cal O}_{\pi K}$ denoted in Eq.~(\ref{EQ:op_pipi})
creates a $\pi K$ state with total isospin $1/2$.

In our concrete calculation we also calculate the ratios
\begin{equation}
\label{EQ:ratio}
R^X(t) = \frac{ C_{\pi K}^X(0,1,t,t\hspace{-0.06cm}+\hspace{-0.06cm}1) }
{ C_\pi (0,t) C_K(1,t\hspace{-0.06cm}+\hspace{-0.06cm}1) },
\quad  X\hspace{-0.05cm}=D, C, \ {\rm and} \ R ,
\end{equation}
where $C_\pi (0,t)$ and $C_K (1,t+1)$ are
the pion and kaon two-point functions, respectively.

\subsubsection{$\kappa$ sector}
In our previous work~\cite{Fu:2011zz},
we make a detailed procedure to measure
$\langle 0 | \kappa^\dag(t)  \kappa   (0) | 0 \rangle$.
For the light $u$ quark Dirac operator $M_u$
and the $s$ quark Dirac operator $M_s$,
we obtain $\kappa$ correlator~\cite{Fu:2011zz}
\begin{equation}
C_\kappa(t) = \sum_{\bf x}(-1)^x e^{i\mathbf{p}\cdot \mathbf{x}}
\langle
\mbox{Tr}[M^{-1}_u({\bf x},t;0,0)M^{-1^\dag}_s( {\bf x},t;0,0)]
\rangle ,
\label{EQ_kappa}
\end{equation}
with given momentum $\mathbf{p}$.

For staggered quarks, the meson propagators have the generic single-particle form,
\begin{equation}
\label{sfits:ch7}
{\cal C}(t) =
\sum_i A_i e^{-m_i t} + \sum_i A_i^{\prime}(-1)^t e^{-m_i^\prime t}  +(t \rightarrow N_t-t),
\end{equation}
where the oscillating terms correspond to a particle with opposite parity.
For the $\kappa$ meson correlator,
we consider only one mass with each parity in Eq.~(\ref{sfits:ch7}),
and the $\kappa$ correlator was fit to
\begin{equation}
\label{eq:kfit}
  C_{\kappa}(t)  = b_{\kappa}e^{-m_{\kappa}t} +
  b_{K_A}(-)^t e^{-M_{K_A}t} + (t \rightarrow N_t-t),
\end{equation}
where the $b_{K_A}$ and $b_{\kappa}$ are two overlap factors.

\subsubsection{ Off-diagonal sector}
To avoid the complicated Fierz rearrangement of the quark lines~\cite{Fukugita:1994ve},
we select $t_1 =0, t_2=1$, and $t_3=t$
for $\pi K \to \kappa$ three-point function,
and choose $t_1 =0, t_2=t$, and $t_3=t+1$
for $\kappa \to \pi K$ three-point function.
\begin{figure}[th]
\begin{center}
\includegraphics[width=6cm,clip]{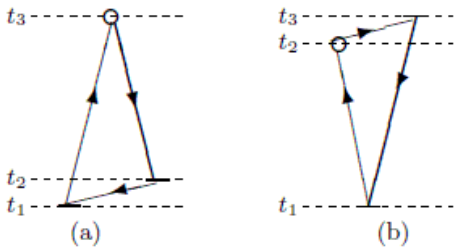}
\end{center}
\caption{ \label{fig:3diagram}
Diagrams contributing to $\pi K \to \kappa$ and
$\kappa \to \pi K$ three-point functions.
(a) Quark contractions of $\pi K \to \kappa$ and
(b) Quark contractions of $\kappa \to \pi K$. 
Short bars show the wall sources.
Open circles are sinks for local pion operator.
The thicker lines indicate the strange quark lines.
}
\end{figure}

The quark line diagrams contributing to $\kappa \to \pi K$ and
$\pi K \to \kappa$ three-point function are plotted
in Fig.~\ref{fig:3diagram}(a) and Fig.\ref{fig:3diagram}(b), respectively.
For the nonzero momentum, we used a $u$ quark source with $1$,
and a strange quark source with $e^{i{\mathbf{p}}\cdot{\mathbf{x}}}$
on each site for kaon  creation operator.
We can express $\pi K \to \kappa$  three-point function 
in terms of the quark propagators $G$ as
\begin{eqnarray}
\label{eq:dcr3}
\hspace{-0.2cm}C_{\pi K \to \kappa} ({\mathbf p},t_3,t_2,t_1) &=& \nonumber \\
&&\hspace{-4.0cm}\sum_{ {\mathbf{x}}_3, {\mathbf{x}}_1} e^{ i {\mathbf p} \cdot {\mathbf{x}}_3 }
\langle \mbox{Tr}
[\gamma_5 G_{t_1}({\mathbf{x}}_3, t_3) 
        G_{t_2}^{\dag}({\mathbf{x}}_3, t_3)
 \gamma_5 G_{t_2}^{\dag}({\mathbf{x}}_1, t_1) ] ,
\end{eqnarray}
where trace is over the color index; the $\gamma_5$  matrix
are utilized as an interpolating field for the pseudoscalars.

\subsection{ Extraction of energies }
\label{SubSec:Extraction of energies}
To reliably obtain two lowest energy eigenstates,
we employ the variational method~\cite{Luscher:1990ck}
and construct a ratio of the correlation function
matrices as
\begin{equation}
M(t,t_R) = C(t) \, C^{-1}(t_R) ,
\label{eq:M_def}
\end{equation}
with some reference time slice $t_R$~\cite{method_diag}.
Two lowest energy levels can be obtained
by a correct fit to two eigenvalues $\lambda_n (t,t_R)$ ($n=1,2$) of the matrix $M(t,t_R)$.
Since we work on the staggered fermion, 
$\lambda_n (t,t_R)$ ($n=1,2$) are  behaved as~\cite{Fu:2011wc}
\begin{eqnarray}
\label{Eq:asy}
\lambda_2 (t, t_R) &=&  A_n \cosh(-E_n(t-T/2)) \nonumber \\
&+& (-1)^t B_n \cosh(-E_n^{\prime}(t-T/2)) \,,
\end{eqnarray}
which explicitly contains an oscillating term,
for a large $t$, which mean $0 \ll t_R < t \ll T/2$ to
suppress the excited states and the wrap-around contaminations~\cite{Feng:2010es}.
Here we suppose the non-degenerate eigenvalues $\lambda_1(t,t_R) > \lambda_2(t,t_R)$.

\section{Lattice calculation}
\label{sec:latticeCal}

We used the MILC lattice ensemble with the $N_f=2+1$ dynamical flavors of the Asqtad-improved
staggered dynamical fermions, see the more details
in Refs.~\cite{Bernard:2010fr,Bazavov:2009bb}.
We measured $\pi K$ four-point functions on the $0.15$~fm MILC lattice ensemble
of $360$ $16^3 \times 48$ gauge configurations
with bare quark masses $am_{ud} = 0.0097$
and $am_s = 0.0484$ and bare gauge coupling $10/g^2 = 6.572$,
which has a physical volume approximately $2.5$~fm, 
and the inverse lattice spacing $a^{-1}=1.358^{+35}_{-13}$ GeV~\cite{Bernard:2010fr,Bazavov:2009bb}.
The mass of the dynamical strange quark is near to its physical value,
and the masses of the $u$ and $d$ quarks have the same mass.

We use the standard conjugate gradient method to get
the required matrix element of the inverse fermion matrix.
The correlator $C_{11}(t)$ is calculated by
\begin{eqnarray}
 C_{11}(t) &=&
\frac{1}{T}\sum_{t_s}\left\langle
\left(\pi K\right)(t+t_s)\left(\pi K\right)^\dag(t_s)\right\rangle .
\end{eqnarray}
After averaging the correlator over all $48$ possible values, 
the statistics are greatly improved.

We calculate off-diagonal correlator $C_{21}(t)$ through
\begin{equation}
C_{21}(t) = \left\langle\kappa(t)(\pi K)^\dag(0)\right\rangle=
\frac{1}{T}\sum_{t_s}
\left\langle\kappa(t+t_s)(\pi K)^\dag(t_s)\right\rangle ,
\end{equation}
where, again, we sum the correlator over all time slices $t_s$ and average it.
Exploiting the relation $C_{12}(t)=C_{21}^\ast(t)$,
we can obtain the off-diagonal matrix element $C_{12}$.
In the following analysis we replace matrix element $C_{12}$
with the matrix element $C_{21}$.

We construct the $\kappa$-correlator
\begin{equation}
C_{22}(t)=\left\langle\kappa^\dag(t+t_s) \kappa(t_s)\right\rangle ,
\end{equation}
where, also, we sum the correlator over all time slices $t_s$ and average it.

In this work, we also measure two-point correlator
for the pion and kaon, namely,
\begin{eqnarray}
G_\pi(t;{\mathbf p}) &=&
\langle 0|\pi^\dag ({\mathbf 0},t) \pi({\mathbf 0},t_S) |0\rangle \,, \nonumber \\
G_K(t;  {\mathbf p}) &=&
\langle 0|  K^\dag ({\mathbf p},t)   K({\mathbf p},t_S) |0\rangle \,,
\label{eq:Gpi}
\end{eqnarray}
where the $G_\pi (t;{\mathbf 0})$ is correlation function
for the pion with zero momentum,
and the $G_K (t;{\mathbf p})$ is correlation function
for the kaon with the momentum ${\mathbf p}$.

\section{Simulation results }
\label{Sec:Results}
\subsection{Time correlation function}
In Fig.~\ref{fig:ratio} the individual ratios are displayed as the functions of $t$.
The values of the direct amplitude $R^D$ is quite close to unity,
indicating a weak interaction in this channel.
The crossed amplitude increases linearly, suggesting a repulsion in this channel.
After an initial increase up to $t \sim 4$, the rectangular amplitude
demonstrates a roughly linear decrease up to $t \sim 15$, 
implying an attractive force.
These features are what we wanted~\cite{Bernard:1990kw,Sharpe:1992pp}.
\begin{figure}[h]
\begin{center}
\includegraphics[width=8cm,clip]{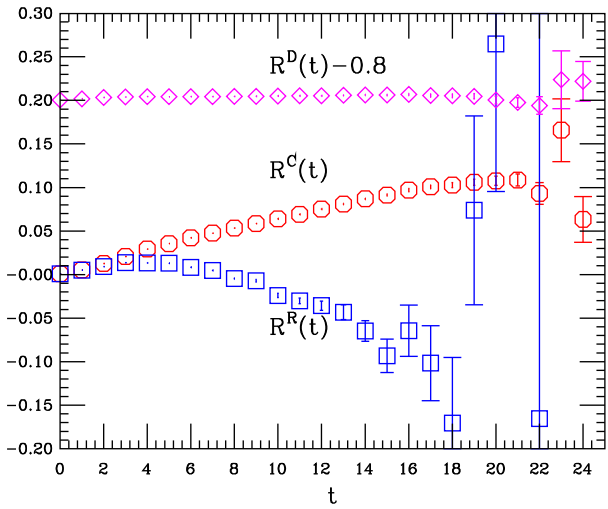}
\end{center}
\caption{(color online).
Individual amplitude ratios $R^X(t)$ for $\pi K$ four-point function
calculated by wall source as functions of $t$.
Direct diagram shifted by $0.8$ (diamonds), crossed diagram (octagons)
and rectangular (squares) diagrams.
\label{fig:ratio}
}
\end{figure}

In Fig.~\ref{fig:G_t}, we show the real parts of the diagonal components
($\pi K\to\pi K$ and $\kappa\to\kappa$)
and the real part of the off-diagonal component $\pi K\to\kappa$
of the correlation function $C(t)$ in Eq.~(\ref{eq:CorrMat}).
\begin{figure}[h]
\begin{center}
\includegraphics[width=8.0cm]{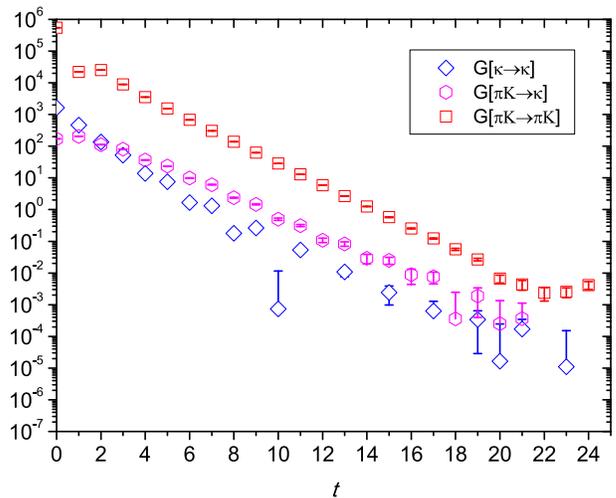}
\end{center}
\caption{ \label{fig:G_t}
(color online). Real part of the diagonal components
($\pi K\to\pi K$ and $\kappa\to\kappa$)
and the real part of the off-diagonal component $\pi K\to\kappa$
of the time correlation function $C(t)$.
Occasional points with negative central values
for the off-diagonal component $\pi K\to\kappa$ are not plotted.
}
\end{figure}
We calculate two eigenvalues $\lambda_n(t,t_R)$ ($n=1,2$)
for the matrix $M(t,t_R)$ in Eq.~(\ref{eq:M_def}) with the reference time $t_R = 6$.
In Figure~\ref{fig:Lambda_t}
we display our lattice results for $\lambda_n(t, t_R) (n = 1, 2)$
as a function of time $t$.
By fitting $\lambda_n(t, t_R) (n = 1, 2)$,
we then extract the energies which are used to calculate the scattering phase.

\begin{figure}[h]
\begin{center}
\includegraphics[width=8.0cm]{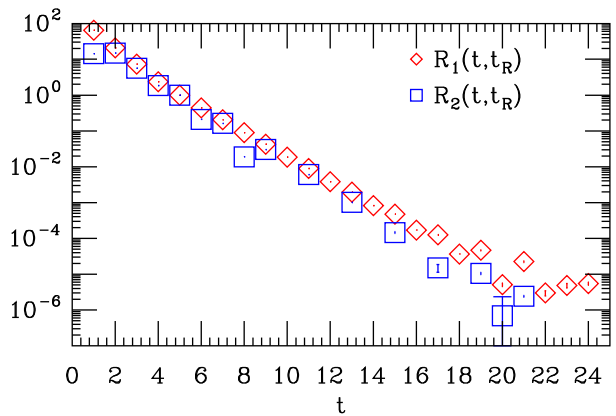}
\end{center}
\caption{\label{fig:Lambda_t}
(color online). The eigenvalues $\lambda_1(t,tR)$ and $\lambda_2(t, tR)$
Occasional points with negative central values
for the correlator $\lambda_2(t, tR)$ are not displayed.
}
\end{figure}

We must consider the contaminations from  the excited states
and the warp-around contributions~\cite{Feng:2010es}.
By denoting a fitting range $[t_{\mathrm{min}}, t_{\mathrm{max}}]$
and varying the values of the $t_{\mathrm{min}}$ and $t_{\mathrm{max}}$,
we can obtain the energies reliably~\cite{Feng:2010es}.
In our concrete fitting, we set $t_\mathrm{min} = t_R+1$ and
increase the reference time slice $t_R$ to
suppress the excited state contaminations~\cite{Feng:2010es}.
Moreover, we select $t_\mathrm{max}$ to be 
away from $t=T/2$ to avert the warp-around effects~\cite{Feng:2010es}.
The fitting parameters:
$t_R$, $t_{\mathrm{min}}$ and $t_{\mathrm{max}}$
are tabulated in Table~\ref{tab:fitting_results}.
The fitted values for $\overline{E}_n$ ($n=1,2$) together 
with the fit quality $\chi^2/\mathrm{dof}$  are also given
in Table~{\ref{tab:fitting_results}}.
\begin{table}[h!]
\caption{\label{tab:fitting_results}
The fitted values of the energy eigenvalues $\overline{E}_n$ ($n=1,2$) in lattice units.
Here we also show the reference time $t_R$, the fitting range, $t_{\mathrm{min}}$ and $t_{\mathrm{max}}$, the fit quality $\chi^2/\mathrm{dof}$.
}
\begin{ruledtabular}
\begin{tabular}{cccccc}
n&$t_R$&$t_{\mathrm{min}}$&$t_{\mathrm{max}}$
& $a\overline{E}_n$ &$\chi^2/\mathrm{dof}$ \\
\hline
$1$  &  $5$  &  $6$ & $14$  & $0.7834(18)$ & $12.2/8$ \\
$2$  &  $5$  &  $6$ & $14$  & $0.9144(37)$ & $6.44/5$ \\
\end{tabular}
\end{ruledtabular}
\end{table}

The energy of the free pion and kaon $E_1$
is calculated from the mass $m_\pi$ and energy $E_K$
obtained by a single exponential fit to
$G_\pi (t;{\mathbf 0})$ and $G_K(t;{\mathbf p})$
in Eq.~(\ref{eq:Gpi}), as $E_1 = m_\pi + E_K$,
which  is also listed in Table~\ref{table:TanDel}.
\begin{table}[h]
\caption{ \label{table:TanDel}
The energy of $\pi K$ system $\overline{E}_n$
and the $s$-wave scattering phase shift $\delta_0$.
$E_1$ is the energy of the free pion and kaon.
$\overline{E}_n$ is obtained from the fitting.
The invariant mass $\sqrt{s}$, the scattering momentum $k$ and the phase $\delta_0$
calculated with relations
(\ref{eq:Disp_Two_Cont_s},\,\ref{eq:Disp_Two_Cont_k}) in the continuum
are called {\it Cont}, and those obtained with the relations
(\ref{eq:Disp_Two_Lat_s},\,\ref{eq:Disp_Two_Lat_k}) on the lattice
are called {\it Lat}. The momentum $k_0$ is calculated by
$k_0^2 = 1/4\times(\sqrt{s} + (m_\pi^2 - m_K^2)/\sqrt{s})^2 - m_\pi^2$.
All values with the mass dimension are in lattice units.
}
\begin{ruledtabular}
\begin{tabular}{ l l l l l }
       & \multicolumn{1}{c}{ $n=1$ } &
       & \multicolumn{1}{c}{ $n=2$ } &  \\
\hline
$E_n$          &  $ 0.8039(15) $  &
                 & \multicolumn{1}{c}{-----} &  \\
$\overline{E}_n$ &  $ 0.7834(18) $  &
                 &  $ 0.9144(37) $      \\
\hline
& \multicolumn{1}{c}{Cont} & \multicolumn{1}{c}{Lat}
& \multicolumn{1}{c}{Cont} & \multicolumn{1}{c}{Lat}  \\
$\sqrt{s}$       &  $ 0.6779(21)$    &   $ 0.6888(21) $
                 &  $ 0.8258(41)$    &   $ 0.8377(41) $  \\
$k^2$            &  $ 0.0112(7) $    &   $ 0.0150(7)  $
                 &  $ 0.0652(16)$    &   $ 0.0716(18) $  \\
$k_0^2$  & \multicolumn{1}{c}{-----} &   $ 0.0148(7)  $
         & \multicolumn{1}{c}{-----} &   $ 0.0700(16) $  \\
$\tan\delta$     &  $ 0.871(34)$     &   $ 0.591(22)  $
                 &  $-1.202(43)$     &   $-1.495(55)  $  \\
$\sin^2\delta$   &  $ 0.431(19)$     &   $ 0.259(14) $
                 &  $ 0.591(17)$     &   $ 0.691(16)  $  \\
\end{tabular}
\end{ruledtabular}
\end{table}
In Table~\ref{table:Disp} we show the mass $m$ and the energy $E$
of the pion and $\kappa$ meson
with momentum ${\mathbf p}=(2\pi/L){\mathbf e}_3$,
calculated from the corresponding time correlation functions~\cite{Fu:2011zz}.
\begin{table}[h]
\caption{\label{table:Disp}
Mass $m$ of the $\pi$, $K$, and $\kappa$ meson,
and energy $E$ of the $K$ and $\kappa$ meson
with momentum ${\mathbf p} = (2\pi/L) {\mathbf e}_3$,
extracted from the correlation function.
}
\begin{ruledtabular}
\begin{tabular}{ c l l c }
     & $\pi$           & K  &  $\kappa$  \\
\hline
$am$  &  $ 0.2459(3) $  & $0.3962(2) $ & $0.758(33)$  \\
$aE$  &                 & $0.5580(14)$ & $0.892(22)$  \\
\end{tabular}
\end{ruledtabular}
\end{table}

\subsection{Lattice discretization effects }
\label{SubSec: Effect of finite lattice spacing }

We should consider the important discretization error
in the $\pi K$ system's Rummukainen-Gottlieb formula (\ref{eq:Luscher_MF}).
It stems from the Lorentz transformation
from the moving frame to the center-of-mass frame
using the Lorentz symmetry in the continuum limit
\begin{eqnarray}
\label{eq:Disp_Two_Cont_s}
\sqrt{s}  &=& \sqrt{ E_{MF}^2 - p^2 }  \,, \\
\label{eq:Disp_Two_Cont_k}
k^2  &=& \frac{1}{4}
\left( \sqrt{s} + \frac{m_\pi^2 - m_K^2}{\sqrt{s}} \right)^2 - m_\pi^2 \,,
\end{eqnarray}
for the invariant mass $\sqrt{s}$,
the energy in moving frame $E_{MF}$ and the momentum $k$.
However, on the lattice, the discretization effects
explicitly break Lorentz symmetry and
Eqs.~(\ref{eq:Disp_Two_Cont_s},\,\ref{eq:Disp_Two_Cont_k})
is only valid up to the discretization errors.
Therefore, the definitions of $\sqrt{s}$ and $k$
contain the similar discretization errors.

These systematic errors have been studied
by Rummukainen and Gottlieb~\cite{Rummukainen:1995vs}, and they
suggest the lattice modified relations.
Following their suggestion,
we recommended the invariant mass $\sqrt{s}$ and the momentum $k$
from the energy in moving frame $E_{MF}$ for $\pi K$ system as
\begin{eqnarray}
\label{eq:Disp_Two_Lat_s}
\cosh( \sqrt{s} ) \hspace{-0.1cm} &=& \hspace{-0.1cm}
\cosh(E_{MF}) - 2\sin^2(p/2)  \,, \\
2\sin^2 (k/2)     \hspace{-0.1cm} &=& \hspace{-0.1cm}
\cosh\left( \frac{\sqrt{s}}{2} \hspace{-0.1cm}+\hspace{-0.1cm}
            \frac{m_\pi^2 \hspace{-0.1cm}-\hspace{-0.1cm} m_K^2}{2\sqrt{s}} \right)
\hspace{-0.1cm}-\hspace{-0.1cm} \cosh(m_\pi) \,,
\label{eq:Disp_Two_Lat_k}
\end{eqnarray}
and then scattering phase shift was obtained by
plugging the momentum $k$ into
the formula in Eq.~(\ref{eq:Luscher_MF}).

To understand these discretization effects,
we calculate the phase shifts from  the energy momentum relations
both in the continuum (\ref{eq:Disp_Two_Cont_s},\,\ref{eq:Disp_Two_Cont_k})
and on the lattice (\ref{eq:Disp_Two_Lat_s},\,\ref{eq:Disp_Two_Lat_k}).
We call the difference coming from
two choices as the discretization error.
These results are listed in Table~\ref{table:TanDel} 
along with the invariant mass $\sqrt{s}$ and the scattering momentum $k$.

\subsection{Extraction of resonance parameters}
\label{SubSec:Scattering Phase Shift and Decay Width }
From Table~\ref{table:TanDel},
we note that the considerable differences due to 
the choices of the energy-momentum relations
are observed in $\sqrt{s}$ and $k$.
Moreover, the difference for the scattering phase $\delta_0$ is
significantly larger than the corresponding statistical errors.
These are also shown in Fig.~\ref{fig:sin2Del},
where the phase shift $\sin^2 \delta_0$ is also drawn.
In Table~\ref{table:TanDel} we watch that
the sign of the scattering phase $\delta$
at $\sqrt{s}< m_\kappa$ ($am_\kappa = 0.758(33)$)
is positive, and that at $\sqrt{s}> m_\kappa$ is negative.
These features confirm a resonance at a mass around $\kappa$ mass $m_\kappa$.

To avert the direct measurement of the decay width~\cite{Aoki:2007rd},
we adopt an alternative approach.
As we discussed in Section~\ref{SubSec:Scattering_Phase},
we parameterize the resonant characteristic of
the $s$-wave phase shift $\delta_0$ by means of
the effective $\kappa\to\pi K$ coupling constant $g_{\kappa\pi K}$, namely,
\begin{equation}
\tan\delta_0 = \frac{g_{\kappa\pi K}^2}{8\pi}\frac{k}{\sqrt{s}(M_R^2-s)} \,,
\label{eq:tanDel_g}
\end{equation}
where $M_R$ is the resonance mass.
The equation~(\ref{eq:tanDel_g}) allows us
to solve for two unknown parameters: $g_{\kappa\pi K}$, $M_R$,
if we assume that the coupling constant $g_{\kappa\pi K}$
varies  quite slowly as the quark mass changes~\cite{Nebreda:2010wv}.
From Table~\ref{table:TanDel}, we can notice 
that the difference between $k$ and $k_0$  is not remarkable.
In practice, we employ the scattering momentum $k_0$ 
when applying Eq.~(\ref{eq:tanDel_g}).

The lattice results of the coupling constant $g_{\kappa\pi K}$
and the resonance mass $M_R$ solved by Eq.~(\ref{eq:tanDel_g}) are
\begin{eqnarray}
g_{\kappa\pi K}  &=&  4.85(86) \, {\rm GeV} \,,  \cr
M_R              &=&  0.742(21)             \,,  \cr
M_R / m_\kappa   &=&  0.980(51)             \,,
\label{eq:FinalR_Cont}
\end{eqnarray}
where we use the expressions (\ref{eq:Disp_Two_Cont_s},\,\ref{eq:Disp_Two_Cont_k})
in the continuum,
and the $\kappa$ meson mass $m_\kappa$ is quoted from
our previous study~\cite{Fu:2011zz}.
If employing the expressions
(\ref{eq:Disp_Two_Lat_s},\,\ref{eq:Disp_Two_Lat_k}) on the lattice,
we have
\begin{eqnarray}
g_{\kappa\pi K}  &=&  4.54(76) \, {\rm GeV} \,,   \cr
M_R              &=&  0.779(27)             \,,   \cr
M_R / m_\kappa   &=&  1.028(58)             \,.
\label{eq:FinalR_Lat}
\end{eqnarray}
This value of coupling constant $g_{\kappa\pi K}$ is in agreement
with  $g_{\kappa\pi K} = 4.94(7) \, {\rm GeV}$  obtained in Ref.~\cite{Oller:2003vf}.

\begin{figure}[h]
\begin{center}
\includegraphics[width=8.0cm]{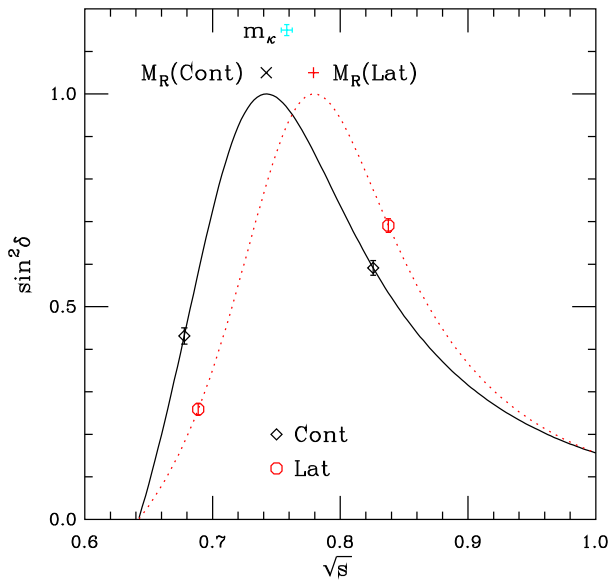}
\end{center}
\caption{\label{fig:sin2Del} (color online).
The scattering phase shift $\sin^2\delta_0$, positions of
$\kappa$ mass $m_\kappa$ and resonance mass $M_R$.
{\bf Cont} refer to the results
achieved with the relations
in the continuum (\ref{eq:Disp_Two_Cont_s},\,\ref{eq:Disp_Two_Cont_k})
and {\bf Lat} to those with the relations
on the lattice (\ref{eq:Disp_Two_Lat_s},\,\ref{eq:Disp_Two_Lat_k}).
Two lines are obtained by Eq.~(\ref{eq:tanDel_g})
with parameters $g_{\kappa\pi K}$ and $M_R$
given in Eq.~(\ref{eq:FinalR_Cont}) and Eq.~(\ref{eq:FinalR_Lat}), respectively.
The abscissa is in lattice units.
}
\end{figure}

In Figure~\ref{fig:sin2Del} we display the curves for $\sin^2\delta_0$
obtained by Eq.~(\ref{eq:tanDel_g})
with the coupling constant $g_{\kappa\pi K}$ and
the resonance mass $M_R$ given in Eq.~(\ref{eq:FinalR_Cont})
and Eq.~(\ref{eq:FinalR_Lat}), respectively.
The position at $\sin^2\delta_0=1$ which stand for
$M_R$ is also marked in Figure~\ref{fig:sin2Del}
for two cases
(black cross and red plus for the continuum and lattice case, respectively).
For visualized comparison,
we also draw the kappa mass $m_\kappa$ (fancy cyan plus).
We can note that $M_R$ is in reasonable agreement with
the kappa mass $m_\kappa$.

Assuming that the dependence of $g_{\kappa\pi K}$ on quark mass
is small~\cite{Nebreda:2010wv},
we can roughly estimate the $\kappa$ meson decay width
at the physical quark mass as
\begin{equation}
\Gamma^{\rm phy} =
\frac{g_{\kappa\pi K}^2}{8\pi}
\frac{k^{\rm phy}}{(m_\kappa^{\rm phy})^2} \,,
\end{equation}
where $m_\kappa^{\rm phy}=849$ MeV is
the physical $\kappa$ meson mass,
which we take from the most recent BESII experiment~\cite{Ablikim:2010kd},
and momentum $k^{\rm phy}$ is calculated by
$$
(k^{\rm phy})^2 = \frac{1}{4}
\left(  m_\kappa^{\rm phy} + \frac{(m_\pi^{\rm phy})^2 -(m_K^{\rm phy})^2}{ m_\kappa^{\rm phy}} \right)^2 -
(m_\pi^{\rm phy})^2  \,,
$$
where $m_\pi^{\rm phy}$ is physical pion mass
($m_\pi^{\rm phy} = 140$ MeV)~\cite{Nakamura:2010zzi},
and $m_K^{\rm phy}$ is physical kaon mass
($m_K^{  \rm phy} = 494$ MeV)~\cite{Nakamura:2010zzi}.
This produces
\begin{equation}
\label{eq:FinalR_Gamm_Cont}
\Gamma^{\rm phy}  = (335 \pm 118) \, {\rm MeV} \,
\end{equation}
where we use the data given in Eq.~(\ref{eq:FinalR_Cont}), and
\begin{equation}
  \Gamma^{\rm phy} = (293  \pm 101)  \, {\rm MeV}
\label{eq:FinalR_Gamm_Lat}
\end{equation}
where we employ the data given in Eq.~(\ref{eq:FinalR_Lat}).
These lattice estimate is fairly near to
the corresponding most recent BESII experimental data~\cite{Ablikim:2010kd}
for the $\kappa \to \pi K$ decay width, $\Gamma_\kappa = 512 \pm 80$~MeV.
We can observe that the difference stemming from
our two chosen energy-momentum relations
is comparable with the statistical error.
This is a quite inspiring result,
considering that we assume the coupling constant is independent of the quark mass,
conduct a long chiral extrapolation, etc.

\section{Conclusions and outlooks}
\label{Sec:Conclusions}
In this work, we have performed out a direct lattice QCD calculation
of the $s$-wave $\pi K$ scattering phase for the $I = 1/2$ channel
near the $\kappa$-meson resonance region in the moving frame,
for the MILC ``medium'' coarse ($a=0.15$ fm) lattice
ensemble in the presence of $2+1$ flavors of the Asqtad
improved staggered dynamical sea quarks.
We employed the same technique in Ref.~\cite{Kuramashi:1993ka}
to calculate all three diagrams categorized in Ref.~\cite{Nagata:2008wk},
and obtained the pretty good signals.
We have demonstrated that the calculation of the $s$-wave scattering phase shift
for the $I=1/2$ $\pi K$ system and then estimation of
the decay width of $\kappa$ meson are practicable.
The scattering phase data shows the presence of
a resonance at a mass around the $\kappa$ meson mass
obtained in~\cite{Fu:2011zz}.
This resonance can be reasonably identified with the $\kappa$ meson.
Moreover, our extracted the $\kappa$ meson decay  
is fairly compared with the $\kappa$ meson decay width
of the most recent BESII experimental measurement.

However, we realize some important issues which
should be cleared in the future works.
One is to reduce the discretization errors,
which we show in previous section are comparable with statistical errors.
An simple way to solve this problem is to
use a lattice configuration closer to the continuum limit.
Another important topics is to suppress the contaminations of
the $s$-wave scattering phase from the $p$-wave scattering phase or higher,
which we are preliminarily discussed for $\pi K$ system
in Ref.~\cite{Fu:2011xz}.

We adopted the effective range formula,
which allows us to use the effective $\kappa \to \pi K$ coupling constant
$g_{\kappa\pi K}$ to extrapolate
from our lattice simulation point $(m_\pi+m_K)/m_\kappa = 0.8$
to the physical point $(m_\pi+m_K)/m_\kappa = 0.73$,
supposing that $g_{\kappa\pi K}$ is independent of the quark mass.
This is just a crude estimation, a more direct evaluation of the decay width is highly desirable.
the decay width can be estimated directly
from the energy dependence of the phase shift data by fitting the BWRF
if we have the simulation data which have several energy
near the resonance mass as it was done for the calculation of the $\rho$
meson decay width in Ref.~\cite{Feng:2010es}.

Although a precise determination of the $\kappa$ resonance parameters
on the lattice is absolutely a big challenge,
our preliminary work reported here can serve as
stepping out the solid first step in an attempt to
study the strong $\kappa$ decay directly from lattice QCD
in a conceptual manner.

\section*{Acknowledgments}
We thank Carleton DeTar for kindly providing us the MILC gauge configurations
used for this work and the fitting software to analyze the lattice simulation data.
We should thank Naruhito Ishizuka for the constructive help
about group symmetry.
We are indebted to MILC Collaboration
for using the Asqtad lattice ensemble and MILC codes.
We are grateful to Hou Qing for his support.
The computations for this work were carried out at AMAX,
CENTOS and HP workstations
in the Institute of Nuclear Science and Technology, Sichuan University.

\appendix
\section{The calculation method of zeta function }
\label{appe:zeta}
Here we briefly discuss one useful method
for the numerical evaluation of the zeta function $\mathcal{Z}_{00}(s;q^2)$
defined in Eq.~(\ref{eq:Zeta00_CM}) in the center-of-mass system
or Eq.~(\ref{eq:Luscher_MF}) in the moving frame
for any value of $q^2$ (i.e., negative or positive).
Here we follow the original derivations and notations
in Ref.~\cite{Yamazaki:2004qb}.

The definition of zeta function $\mathcal{Z}_{00}^{\mathbf d}(s;q^2)$
in Eq.~(\ref{eq:Zeta00_CM}) is
\begin{equation}
\sqrt{ 4\pi } \cdot \mathcal{Z}^{ \mathbf d }_{ 00 } ( s ; q^2 ) =
\sum_{ {\mathbf r} \in P^{\mathbf d} } \frac{1 }{( r^2 - q^2 )^s } \,,
\label{eq:Z00d_appendix}
\end{equation}
where the summation for ${\mathbf r}$ is carried out over the set
\begin{equation}
\label{eq:app:pset}
P_{\mathbf d} = \left\{ {\mathbf r} \left|  {\mathbf r} =
\vec{\gamma}^{-1}
\left({\mathbf n}+\frac{\alpha}{2} {\mathbf d} \right) \, \quad
{\mathbf n}\in \mathbb{Z}^3 \right. \right\} \,,
\end{equation}
where
$\displaystyle
\alpha = 1 + \frac{m_K^2-m_\pi^2}{E_{CM}^2} \,.
$
The operation $\hat{ \gamma }^{ -1 }$ is denoted in Eq.~(\ref{la:gamma}).
Without loss of generosity,
we consider that the value $q^2$ can be a positive or negative.

First we consider the case of $q^2 > 0$,
and separate the summation in $\mathcal{Z}_{00}^{\mathbf d}( s ; q^2 )$ into two parts as
\begin{equation}
\sum_{{\mathbf r} \in P_{\mathbf d}} \frac{1}{( r^2 - q^2 )^{s}} =
\sum_{r^2 < q^2}\frac{1}{(r^2-q^2)^{s}} +
\sum_{r^2 > q^2}\frac{1}{(r^2-q^2)^{s}} \,,
\label{eq:app_zeta_1}
\end{equation}
where the summation over ${\mathbf r }$ is carried out
with ${\mathbf r }\in P^{ \mathbf d }$ in Eq.~(\ref{eq:app:pset}).
The second term can be written in an integral form,
\begin{widetext}
\begin{eqnarray}
\sum_{ r^2 > q^2 } \frac{1}{( r^2 - q^2 )^{s}}
&=&
\frac{1}{ \Gamma(s) } \sum_{r^2 > q^2 }
\left[ \int_0^1        {\rm d}t \ t^{s-1} e^{ - t ( r^2 - q^2) } +
       \int_1^{\infty} {\rm d}t \ t^{s-1} e^{ - t ( r^2 - q^2) } \right] \cr
&=&
\frac{1}{\Gamma(s)}
\int_0^1 {\rm d}t  t^{s-1} e^{q^2 t} \sum_{\mathbf r} e^{-r^2 t}
\hspace{-0.1cm}-\hspace{-0.1cm}
\sum_{ r^2 < q^2 } \frac{1}{( r^2 - q^2 )^{s}}
\hspace{-0.1cm}+\hspace{-0.1cm}
\sum_{\mathbf r} \frac{ e^{ -( r^2 - q^2 ) } }{ (r^2 - q^2)^s }.
\label{eq:app_zeta_2}
\end{eqnarray}
The second term neatly cancels out the first term in Eq.~(\ref{eq:app_zeta_1}).
Next we rewrite the first term in Eq.~(\ref{eq:app_zeta_2})
by the Poisson's resummation formula as
\begin{equation}
\hspace{-0.01cm}\frac{1}{\Gamma(s)} \int_0^1  {\rm d}t  t^{s-1} e^{ t q^2 }
\sum_{\mathbf r } e^{-r^2 t} =
\frac{\gamma}{ \Gamma(s) }
\int_0^1  {\rm d}t  t^{s-1} e^{t q^2}
\left( \frac{ \pi }{t} \right)^{\frac{3}{2}}
\sum_{ {\mathbf n } \in \mathbb{Z}^3 }
e^{i \pi \alpha \, \mathbf n \cdot d  }
e^{ \pi^2 (\hat{\gamma}{\mathbf n} )^2/t } \,.
\label{eq:app_zeta_3}
\end{equation}
The divergence at $s=1$ comes from the ${\mathbf n} = {\mathbf 0}$ part
of the integrand. Therefore we divide the integrand into
a divergent part (${\mathbf n} = {\mathbf 0}$) and a finite part (${\mathbf n} \ne {\mathbf 0}$).
The divergent part can be evaluated for $ {\mathrm Re}\,s > 3 / 2 $ as
\begin{equation}
\int_0^1 \ {\rm d}t \ t^{s-1}
e^{q^2 t} \left( \frac{ \pi }{ t } \right)^{ \frac{3}{2} } =
\sum_{l=0}^{ \infty }\frac{ \pi^{3/2}}{s+l-3/2}\frac{q^{2l}}{l!}.
\label{eq:app_zeta_4}
\end{equation}
The right hand side of this equation has a finite value at $s=1$.

After connecting all terms we obtain the representation of
the zeta function  at $s=1$,
\begin{equation}
\sqrt{4\pi} \cdot \mathcal{Z}_{00}^{\mathbf d} (1; q^2) =
\sum_{ \mathbf r } \frac{ e^{ -(r^2-q^2) } }{r^2-q^2}
+ \gamma \int_0^1 \ {\rm d}t \ e^{ t q^2}
\left( \frac{ \pi }{ t } \right)^{ \frac{3}{2} }
\sum_{ {\mathbf n } \in \mathbb{Z}^3 } \!\!\rule{0mm}{1em}^{\prime}
e^{ i \pi \alpha  \, {\mathbf n} \cdot {\mathbf d} }
e^{ \pi^2 ( \hat{ \gamma }{\mathbf n } )^2/t }  +
\gamma \sum_{l=0}^{\infty} \frac{ \pi^{3/2} }{l-1/2}
\frac{ q^{2l} }{ l ! } ,
\label{eq:app_zeta_s=1}
\end{equation}
where $\sum_{ {\mathbf n} \in \mathbb{Z}^3 }^{\prime}$ stands for a summation without ${\mathbf n} = {\mathbf 0}$.

For the case of $q^2 \le 0$, it is not necessary  to divide the summation
in $\mathcal{Z}_{00}(s; q^2)$, and it can be also expressed in an integral form,
\begin{equation}
\sum_{{\mathbf r} \in P_{\mathbf d} } \frac{1}{(r^2-q^2)^{s}} =
\sum_{{\mathbf r} \in P_{\mathbf d} } \frac{ e^{ -(r^2-q^2) } }{ (r^2-q^2)^s }
+
\frac{ \gamma }{ \Gamma ( s ) }
\int_0^1  {\rm d}t \ t^{ s - 1 } e^{t q^2}
\left( \frac{ \pi }{ t } \right)^{ \frac{3}{2} }
\sum_{ {\mathbf n } \in \mathbb{Z}^3}
e^{ i \pi \alpha \, {\mathbf n \cdot d } }
e^{ \pi^2 ( \hat{ \gamma }{\mathbf n} )^2 / t } .
\end{equation}
Following the same procedures, we arrive at the same expression
in Eq.~(\ref{eq:app_zeta_s=1}).
Hence, Eq.~(\ref{eq:app_zeta_s=1}) can be applied for both cases.

Substituting ${\mathbf d} = (1,0,0)$ into Eq.~(\ref{eq:app_zeta_s=1})
we obtain the representation of the zeta function
appeared in Eq.~(\ref{zetafunction_MF})
\begin{equation}
\sqrt{4\pi} \cdot \mathcal{Z}_{ 00 }^{\mathbf d} (1; q^2) =
\sum_{ \mathbf r } \frac{ e^{ -(r^2-q^2) } }{r^2-q^2}
+ \gamma \sum_{ l = 0 }^{ \infty } \frac{ \pi^{ \frac{3}{2}} }{ l - 1 / 2 }
\frac{ q^{2l} }{ l ! }
+
\gamma \int_0^1  {\rm d}t  e^{ t q^2}
\left( \frac{ \pi }{ t } \right)^{ \frac{3}{2} }
\sum_{ {\mathbf n } \in \mathbb{Z}^3} \!\!\rule{0mm}{1em}^{\prime}
\cos( \alpha  \pi \, {\mathbf n} \cdot {\mathbf d} )
e^{ \pi^2 ( \hat{ \gamma }{\mathbf n } )^2 / t } \,,
\end{equation}
where the imaginary parts are neatly canceled out.

Substituting ${\mathbf d} = {\mathbf 0}$ and $\gamma = 1$ into
into Eq.~(\ref{eq:app_zeta_s=1}),
we obtain the representation of the zeta function
in the CM system appeared in Eq.~(\ref{eq:Zeta00_CM})
\begin{equation}
\sqrt{4\pi} \cdot \mathcal{Z}_{00}(1; q^2)  =
\sum_{{\mathbf n } \in \mathbb{Z}^3}
\frac{e^{-(n^2-q^2)}}{ n^2 - q^2} +
\int_0^1  {\rm d}t  e^{t q^2}  \left( \frac{ \pi }{t} \right)^{ \frac{3}{2} }
\sum_{{\mathbf n } \in \mathbb{Z}^3} \!\!\rule{0mm}{1em}^{\prime}
e^{\pi^2 n^2/t} +
\sum_{ l=0}^{\infty}
\frac{ \pi^{3/2} }{l-1/2} \frac{q^{2l}}{l!} \,.
\label{eq:Z_C:s=1}
\end{equation}
\end{widetext}
I also note that, for negative $q^2$, an asymptotic expression
of the zeta function $\mathcal{Z}_{00}( s ; q^2 )$ has been derived in
Ref.~\cite{Davoudi:2011md}.
We numerically compared this representation of the zeta functions
with that of above described representation, and found reasonable agreement.


%

\begin{thebibliography}{9999}
\bibitem{Nakamura:2010zzi} K.~Nakamura {\it et al.} (Particle Data Group),
J. Phys. G {\bf 37}, 075021 (2010).

\bibitem{Bugg:2005xx} D.~V.~Bugg,
Phys. Lett. B {\bf 632}, 471 (2006).

\bibitem{Bugg:2009uk} D.~V.~Bugg,
Phys. Rev.  D {\bf 81}, 014002 (2010).

\bibitem{Aitala:2002kr} M.~Aitala {\it et al.},
Phys. Rev. Lett. {\bf 89}, 121801 (2002).

\bibitem{Ablikim:2010kd} M.~Ablikim {\it et al.},
Phys. Lett. B {\bf 693}, 88 (2010).

\bibitem{Ablikim:2005ni} M.~Ablikim {\it et al.} (BES Collaboration),
Phys. Lett. B {\bf 633}, 681 (2006).

\bibitem{Bai:2003fv} J.~Z.~Bai {\it et al.} (BES Collaboration),
arXiv:hep-ex/0304001.

\bibitem{Alde:1997ri} D.~Alde {\it et al.} (GAMS Collaboration),
Phys. Lett. B {\bf 397}, 350 (1997).

\bibitem{Markushin:1999fg} V.~E.~Markushin and M.~P.~Locher,
Frascati Phys. Ser. {\bf 15}, 229 (1999).

\bibitem{Xiao:2000kx} Z.~Xiao, H.~Q.~Zheng,
Nucl. Phys. A {\bf 695}, 273 (2001).

\bibitem{Cawlfield:2006hm} C.~Cawlfield {\it et al.} (CLEO Collaboration),
Phys. Rev.  D {\bf 74}, 031108 (2006).

\bibitem{Bonvicini:2008jw} G.~Bonvicini {\it et al.} (CLEO Collaboration),
Phys. Rev. D {\bf 78}, 052001 (2008).

\bibitem{Zhou:2006wm} Z.~Y.~Zhou and H.~Q.~Zheng,
Nucl. Phys. A {\bf 775}, 212 (2006).

\bibitem{DescotesGenon:2006uk} S.~Descotes-Genon and B.~Moussallam,
Eur. Phys. J.  C {\bf 48}, 553 (2006).

\bibitem{Ishida:1997wn} S.~Ishida, M.~Ishida, T.~Ishida, K.~Takamatsu and
T.~Tsuru,
Prog. Theor. Phys.  {\bf 98}, 621 (1997).

\bibitem{Bugg:2003kj} D.~V.~Bugg,
Phys. Lett. B {\bf 572}, 1 (2003).

\bibitem{Zheng:2003rw} H.~Q.~Zheng, Z.~Y.~Zhou, G.~Y.~Qin, Z.~Xiao, J.~J.~Wang,
and N.~Wu, Nucl. Phys. A {\bf 733}, 235 (2004).

\bibitem{Pelaez:2004xp} J.~R.~Pelaez,
Mod. Phys. Lett. A  {\bf 19}, 2879 (2004).

\bibitem{Nebreda:2010wv} J.~Nebreda and J.~R.~Pelaez,
Phys. Rev. D {\bf 81}, 054035 (2010).

\bibitem{Fu:2011zz} Z.~-W.~Fu and C.~DeTar,
Chin.\ Phys.\ C {\bf 35}, 1079 (2011).

\bibitem{Fu:2011wc} Z.~Fu,
Phys.\ Rev.\ D {\bf 85}, 074501 (2012).

\bibitem{Bernard:2010fr} C.~Bernard {\it et al.},
Phys. Rev.  D {\bf 83}, 034503 (2011).

\bibitem{Bazavov:2009bb} A.~Bazavov {\it et al.},
Rev. Mod. Phys. {\bf 82}, 1349 (2010).

\bibitem{Luscher:1986pf} M.~Luscher,
Commun. Math. Phys. {\bf 105}, 153 (1986).

\bibitem{Luscher:1991cf} M.~Luscher,
Nuclear Physics B {\bf 354}, 531 (1991).

\bibitem{Lellouch:2001p4241} L.~Lellouch and M.~Luscher,
Commun. Math. Phys. {\bf 219}, 31 (2001).

\bibitem{Luscher:1990ck} M.~Luscher and U.~Wolff,
Nucl. Phys. B {\bf 339}, 222 (1990).

\bibitem{Luscher:1991ux} M.~Luscher,
Nucl. Phys. B {\bf 364}, 237 (1991).

\bibitem{Amsler:2008zzb} C.~Amsler {\it et al.},
Phys. Lett. B {\bf 667}, 1 (2008).

\bibitem{Rummukainen:1995vs} K.~Rummukainen and S.~A.~Gottlieb,
Nucl. Phys. B {\bf 450}, 397 (1995).

\bibitem{Davoudi:2011md} Z.~Davoudi and M.~J.~Savage,
arXiv:1108.5371 [hep-lat].

\bibitem{Fu:2011xz} Z.~Fu,
Phys.\ Rev.\ D {\bf 85}, 014506 (2012).

\bibitem{Sharpe:1992pp} S.~R.~Sharpe, R.~Gupta and G.~W.~Kilcup,
Nucl. Phys. B {\bf 383}, 309 (1992).

\bibitem{Kuramashi:1993ka} Y.~Kuramashi, M.~Fukugita, H.~Mino, M.~Okawa and A.~Ukawa,
Phys. Rev. Lett.  {\bf 71}, 2387 (1993).

\bibitem{Fukugita:1994na}  M.~Fukugita, Y.~Kuramashi, H.~Mino, M.~Okawa and A.~Ukawa,
Phys. Rev. Lett.  {\bf 73}, 2176 (1994).

\bibitem{Fukugita:1994ve} M.~Fukugita, Y.~Kuramashi, M.~Okawa, H.~Mino and A.~Ukawa,
Phys. Rev. D {\bf 52}, 3003 (1995).

\bibitem{Nagata:2008wk} J.~Nagata, S.~Muroya and A.~Nakamura,
Phys. Rev. C {\bf 80}, 045203 (2009).

\bibitem{Sasaki:2010zz} K.~Sasaki, N.~Ishizuka, T.~Yamazaki and M.~Oka,
Prog. Theor. Phys. Suppl. {\bf 186}, 187 (2010).

\bibitem{Gupta:1990mr} R.~Gupta, G.~Guralnik, G.~W.~Kilcup and S.~R.~Sharpe,
Phys. Rev. D {\bf 43}, 2003 (1991).

\bibitem{method_diag} M.~L\"uscher and U.~Wolff,
Nucl. Phys. B {\bf 339}, 222 (1990).

\bibitem{Aoki:2007rd} S.~Aoki {\it et al.} (CP-PACS Collaboration),
Phys.\ Rev.\  {\bf D76}, 094506 (2007).

\bibitem{Bernard:1990kw} V.~Bernard, N.~Kaiser and U.~G.~Meissner,
Nucl. Phys. B {\bf 357}, 129 (1991).

\bibitem{Oller:2003vf} J.~A.~Oller,
Nucl. Phys. A {\bf 727}, 353 (2003).

\bibitem{Feng:2010es} X.~Feng, K.~Jansen and D.~B.~Renner,
Phys. Rev. D {\bf 83}, 094505 (2011).

\bibitem{Yamazaki:2004qb} T.~Yamazaki {\it et al.},
Phys. Rev. D {\bf 70}, 074513 (2004).
\end{thebibliography}
\end{document}